\documentclass[
    reprint,
    preprintnumbers,
    superscriptaddress,
    nofootinbib,
    amsmath,amssymb,
    aps,
    prd,
    floatfix,
    longbibliography]{revtex4-2}
    
    \usepackage{graphicx}
    \usepackage{subfigure}
    \usepackage{subcaption} 
    \usepackage[usenames,dvipsnames]{xcolor} % colour
    \usepackage{mathrsfs} % For \mathscr{}

    \usepackage{pgfplots}
    \usepackage[utf8]{inputenc}
    \usepackage{color}
    \usepackage{hyperref}
    \usepackage[normalem]{ulem} 
    \usepackage{physics}
    \usepackage{enumitem}
    \usepackage{comment}
    \usepackage{bm}
    \usepackage{aas_macros}
    \usepackage[capitalise]{cleveref}

\hypersetup{
  breaklinks = true,
  colorlinks   = true, %Colours links instead of ugly boxes
  urlcolor     = blue, %Colour for external hyperlinks
  linkcolor    = blue, %Colour of internal links
  citecolor   = blue %Colour of citations
}

\usepackage{float}

\newcommand{\Cov}{{\rm Cov}}

\newcommand{\oxford}{Astrophysics, University of Oxford, DWB, Keble Road, Oxford OX1 3RH, United Kingdom}

\newcommand{\splitatcommas}[1]{%
  \begingroup
  \begingroup\lccode`~=`, \lowercase{\endgroup
    \edef~{\mathchar\the\mathcode`, \penalty0 \noexpand\hspace{0pt plus 1em}}%
  }\mathcode`,="8000 #1%
  \endgroup
}

\begin{document}

\title{Matching current observational constraints with nonminimally coupled dark energy}

\author{William J. Wolf}
\email{william.wolf@stx.ox.ac.uk}
\affiliation{\oxford}
\author{Pedro G. Ferreira}
\email{pedro.ferreira@physics.ox.ac.uk}
\affiliation{\oxford}
\author{Carlos Garc\'ia-Garc\'ia}
\email{carlos.garcia-garcia@physics.ox.ac.uk}
\affiliation{\oxford}

\begin{abstract}
We show that a Universe with a nonminimally coupled scalar field can fit current measurements of the expansion rate of the Universe better than the standard $\Lambda$-Cold Dark Matter ($\Lambda$CDM) model or other minimally coupled dark energy models. In particular, the nonminimal coupling in this model allows for the dark energy model to exhibit stable phantom crossing behavior, which seems to be suggested by the constraints on the dark energy equation of state coming from the most recent data. While we find a clear improvement in the goodness of fit for this dark energy model with respect to others that have been considered in the recent literature, using information theoretic criteria, we show that the evidence for it is still inconclusive. 
\end{abstract}

\maketitle

%%%%%%%%%%%%%%%%%%%%%%%%%%%%%%%%%%%%%%%%%%%%%%%%%%%%%%%%%%%%%%
\textit{Introduction.---}The latest Dark Energy Spectroscopic Instrument (DESI) measurements of baryon acoustic oscillations (BAO) \cite{DESI:2024mwx}, alongside data from supernovae (SNe) \cite{Rubin:2023ovl, Scolnic:2021amr, DES:2024tys} and the Cosmic Microwave Background (CMB) \cite{Planck:2018vyg, Planck:2019nip,ACT:2023kun, ACT:2023dou}, seem to be hinting at deviations from the $\Lambda$-Cold Dark Matter ($\Lambda$CDM) paradigm and provide evidence for time-evolving dark energy. There has, understandably, been a tremendous amount of work exploring this evidence (see e.g., \cite{Wolf:2024eph, Ye:2024ywg, Tada:2024znt, Park:2024jns, Shlivko:2024llw, Cortes:2024lgw, DESI:2024kob, Dinda:2024kjf, Carloni:2024zpl, Wang:2024rjd, Mukherjee:2024ryz, Roy:2024kni, Wang:2024dka, Gialamas:2024lyw, Notari:2024rti, Wang:2024sgo, Wang:2024hwd, Giare:2024gpk, Dinda:2024ktd, Jiang:2024xnu, Ghosh:2024kyd, Luongo:2024fww, Alfano:2024jqn, Reboucas:2024smm, Pang:2024qyh, Efstathiou_2024, Bhattacharya:2024hep, RoyChoudhury:2024wri, Arjona:2024dsr, Andriot:2024jsh, DESI:2024aqx, Wang:2024tjd, Berghaus:2024kra, Alestas:2024eic, Carloni:2024rrk, Chan-GyungPark:2024brx, Aboubrahim:2024cyk, Ye:2024zpk, Andriot:2024sif, Chudaykin:2024gol, Colgain:2024ksa, Colgain:2024mtg, Colgain:2024xqj, Mello:2024tor, Sapone:2024ltl, Tang:2024lmo}). A particular thread of the literature concerns whether or not the current data, assuming that the recent DESI results hold up to further scrutiny, implies a phantom crossing in the dark energy evolution.

The discussion has largely centered around the well-known Chevallier-Polarski-Linder (CPL) parameterization \cite{Linder:2002et, Chevallier:2000qy}:
\begin{equation}\label{eq:param}
w(a)\simeq w_0+w_a(1-a),
\end{equation}
where, $w=P_{\rm DE}/\rho_{\rm DE}$ is the equation of state of dark energy, with energy density, $\rho_{\rm DE}$,  pressure, $P_{\rm DE}$, and $a$ is the scale factor of the Universe.
The right-hand side of \cref{eq:param} is an approximation to the left-hand side where $w_0$ gives the value of $w$ today and $w_a$ characterizes its temporal evolution. Current constraints on these parameters seem to place dark energy firmly in the phantom regime in the past, where $w(a)<-1$. But, as has been discussed in \cite{Wolf:2023uno, Cortes:2024lgw, Shlivko:2024llw, Wolf:2024eph}, $w_0$ and $w_a$ are in fact fitting parameters and it can be misleading to extrapolate the parametrised $w(a)$ across the entirety of cosmic history. Indeed, there are  non-phantom quintessence models which appear to be phantom when described in terms of the CPL parameterization simply because the resulting parametrised $w(a)$ crosses $-1$ even though the dark energy model itself never dips below $-1$. Nevertheless, if one restricts one-self to thawing models of quintessence (which never have $w(a)<-1$), \citet{Wolf:2024eph} show that there is inconclusive evidence for them from the current data. 

There have been a few attempts at assessing the evidence for the phantom regime from current data. In particular, model agnostic methods have found some evidence for phantom crossing \cite{Ye:2024ywg, DESI:2024aqx} . In light of their results, \citet{Ye:2024ywg} have proposed a nonminimally coupled model of dark energy --  dubbed {\it Thawing Gravity} (TG) -- as a proof of concept to illustrate that nonminimally coupled models with a stable phantom crossing can offer an improvement in terms of fitting the data over $\Lambda$CDM and many minimally coupled quintessence models. In this letter, we will show that this model, when compared with the minimally coupled thawing quintessence of \citet{Wolf:2024eph} and \citet{Wolf:2023uno}, is similar in terms of its consistency with the data. We will show; however, that by building on the insights of \cite{Wolf:2023uno, Wolf:2024eph, Ye:2024ywg}, it is possible to construct a viable, nonminimally coupled dark energy model that does cover the type of evolution of $w(a)$ being uncovered by current data. Nevertheless we will argue that, using current constraints on the redshift dependence of the expansion rate, the evidence is still inconclusive.

\textit{Dark energy models.---}
Our starting point will be the following action:
% \begin{align}\label{eq:fullaction}
% S_{HTG}&=&\int d^4 x\sqrt{-g}\left[\frac{1}{2}\left(M^2_{\rm P}-\xi\varphi^2\right)R-\frac{1}{2}\partial_\mu\varphi\partial^\mu\varphi \right. \nonumber \\ & &\left. -\left(V_1-V_0e^{-\lambda \varphi/M_{\rm P}}\right)+{\cal L}_M\left(g_{\alpha\beta},\psi_M\right)\right],
% \end{align}
\begin{align}\label{eq:fullaction}
S_{HTG}&=&\int d^4 x\sqrt{-g}\left[\frac{1}{2}\left(M^2_{\rm P}-\xi\varphi^2\right)R-\frac{1}{2}\partial_\mu\varphi\partial^\mu\varphi \right. \nonumber \\ & &\left. -V_1\left(1-V_0e^{-\lambda \varphi/M_{\rm P}}\right)+{\cal L}_M\left(g_{\alpha\beta},\psi_M\right)\right],
\end{align}
where $g_{\alpha\beta}$ is the metric, $g$ its determinant and $R$ its Ricci tensor, $\varphi$ is the quintessence field (the source of dark energy), $\xi$ is the nonminimal coupling parameter, ${\cal L}_M(\cdots)$ is the matter action, and $\psi_M$ is a schematic representation of all the matter fields apart from the quintessence field. This model has a large region of parameter space which is safe from ghost and gradient instabilities, which are known to frequently plague theories of modified gravity \cite{Rubakov:2014jja, Wolf:2019hzy, Sbisa:2014pzo, Bellini:2014fua, ErrastiDiez:2024hfq}. 

There are two crucial ingredients to this model: the nonminimal coupling and the form of the potential. The nonminimal coupling allows for phantom crossing behavior \cite{Nesseris:2006er}, which seems to be implied by the data (notwithstanding the earlier mentioned caveats). The potential proposed in \cref{eq:fullaction} allows for a  $d^2 V/d\varphi^2<0$, i.e., a potential with hilltop or plateau-like features. More generally, potentials with this feature have been widely explored in the context of inflation (see e.g., \cite{Boubekeur:2005zm, Wolf:2024lbf, Kallosh:2019jnl}), and to a somewhat lesser extent dark energy \cite{Wolf:2023uno, Dutta:2008qn, Tada:2024znt, Chiba:2009sj, Shlivko:2024llw, Wolf:2024eph}. This particular form of the potential with a negative exponential resembles some that have previously been studied in various contexts in modeling early Universe physics \cite{Steinhardt:2002ih, Cicoli:2008gp, Ijjas:2019pyf}. In contrast with common potentials like the standard quadratic or exponential models, these potentials are concave down and once the field begins rolling down the potential, the evolution of $\varphi$ can be very rapid. This can translate into sharp, non-linear evolution in $w(a)$ which would suggest that such models lie in the steeper $(w_0, w_a)$ regions favoured by current constraints \cite{Wolf:2023uno, Shlivko:2024llw, Wolf:2024eph}. Thus, we will refer to this model as \textit{Hilltop Thawing Gravity} (HTG).

In order to solve the background cosmological equations of motion and the dynamics for linear perturbations, we have implemented this model in \texttt{hi\_class} \cite{hi_class1,hi_class2,CLASS}. We set $\varphi_{ini}=0$ and $\dot{\varphi}_{ini}\neq 0$ (i.e., setting it to a small non-zero value). We tune $V_1$ to satisfy the equations of motion, while varying the parameters $\lambda$, $V_0$, and $\xi$. Similar to \cite{Ye:2024ywg}, we find that $V_1 \simeq 3 H_0^2 \Omega_{\varphi}$ provides a robust starting point for the numerical solvers. Through this letter, $M_{\rm P}=1$, $V_0$ will be in units of $M^2_{\rm P}(H_0/h)^{2}$, and both $\xi$ and $\lambda$ are dimensionless. 

Additionally, as we shall soon see, it will be instructive to compare the HTG model explored here with the TG model explored in \citet{Ye:2024ywg} and the \textit{Thawing Quintessence} (TQ) model explored in \citet{Wolf:2023uno} and \citet{Wolf:2024eph}. For reference, we reproduce the actions of these models here. TG is given by:
\begin{align}\label{eq:tg_action}
S_{TG}&=&\int d^4 x\sqrt{-g}\left[\frac{1}{2}\left(M^2_{\rm P}-\xi\varphi^2\right)R-\frac{1}{2}\partial_\mu\varphi\partial^\mu\varphi \right. \nonumber \\ & &\left. - V_0e^{-\lambda \varphi/M_{\rm P}}+{\cal L}_M\left(g_{\alpha\beta},\psi_M\right)\right],
\end{align}
where we have kept the nonminimal coupling prescription which enables recent phantom crossing and we have a similar exponential form, but $d^2 V/d\varphi^2>0$ in this model because it is always the case that $V_0>0$ here, which does restrict us to a more limited range of dynamical behavior. Additionally, one can easily see from this expression that we can recover the TG model from the HTG action with suitable parameter choices in \cref{eq:fullaction}. TQ is given by:
\begin{align}\label{eq:tq_action}
S_{TQ}=\int d^4 x\sqrt{-g}&\left[\frac{1}{2}M^2_{\rm P}R-\frac{1}{2}\partial_\mu\varphi\partial^\mu\varphi \right. \nonumber \\ &\left. -\left(V_0 + \frac{1}{2}m^2\varphi^2\right)+{\cal L}_M\left(g_{\alpha\beta},\psi_M\right)\right],
\end{align}
where the nonminimal coupling has been dropped, meaning that this model cannot venture into the phantom regime of dark energy. However, this model can have either $d^2 V/d\varphi^2>0$ or $d^2 V/d\varphi^2<0$ depending on the sign of $m^2$. As detailed in \cite{Wolf:2023uno, Wolf:2024eph}, this allows it to have considerable dynamical freedom.

\textit{Results.---} The CPL parameters $(w_0, w_a)$ are fitting parameters, a form of data compression. Often in the literature, one takes the approach of directly fitting $w(a)$ over a recent range of redshifts to best approximate the evolution of $w(a)$ in terms of the CPL parameters. In practice though, we cannot measure $w(a)$ directly; ($w_0$, $w_a$) are inferred from data that effectively probe $H(z)\equiv {\dot a}/a(z)$ at different redshifts, $1+z=1/a$ .  Assuming \cref{eq:param}, $H$ becomes:
\begin{equation}\label{eq:hfit}
H^2(a)=H_0^2\left[\Omega_{\mathrm{m}} a^{-3}+\left(1-\Omega_{\mathrm{m}}\right) e^{3 w_a (a-1)} a^{-3(1+w_0+w_a)}\right],
\end{equation}
where $\Omega_{\mathrm{m}}$ is the fractional energy density of non-relativistic matter today and $H_0$ is the Hubble rate today. One then finds the best fit (and associated uncertainties) of \cref{eq:hfit} to an appropriate selection of data.

If one wishes to compare a particular model to the data, represented in terms of ($w_0$, $w_a$), one needs to undertake the exact same procedure; one must find the corresponding $H(z)$ for that model, determine the corresponding observables at the same redshifts as the data has been collected, and find the best fit values of ($w_0$, $w_a$) that reproduce the predicted observables for the dark energy model. In \cite{Wolf:2024eph}, an efficient approach to this was proposed, specifically targeting BAO, SNe, and CMB probes used in the DESI analysis \cite{DESI:2024mwx}. We briefly recap the method. The BAO data is given in terms of angular diameter distances; the SNe data can be compressed into measurements of $E(z) \equiv H(z) / H_0$ at different redshifts \cite{Riess:2017lxs}; the CMB data can be compressed into an acoustic scale $\ell_{\mathrm{A}}$ and shift parameter $R\left(z_{*}\right)$ at the photon decoupling redshift $z_{*}$ \cite{Chen:2018dbv}. Using this combination of variables, one can reproduce the latest results from the DESI collaboration \cite{DESI:2024mwx} to an excellent approximation; thus, we can be sure that using this procedure will allow us to compare like for like when determining a dark energy model's representation in the $(w_0, w_a)$ plane (see \cite[Appendix A]{Wolf:2024eph} for more details). In other words, this procedure is essentially determining the prior distribution of this class of theories in the CPL parameter plane. This then allows us to compare the theory directly to the likelihood for $(w_0, w_a)$ derived from the data and determine the extent of the overlap.

Given these observables, we then find the best-fit CPL parameters by minimizing
\begin{equation}\label{chi2}
\chi^2 = \left(\mathcal{O}^{\rm CPL} - \mathcal{O}^{\varphi} \right)^T \left(\frac{\mathcal{O}^{\rm data}}{\mathcal{O}^{\varphi}}\right)^T {\Cov}^{-1} \frac{\mathcal{O}^{\rm data}}{\mathcal{O}^{\varphi}} \left(\mathcal{O}^{\rm CPL} - \mathcal{O}^{\varphi} \right),
\end{equation}
which gives the distance between $(\xi, \lambda, V_0, V_1)$ and ($w_0$, $w_a$). Here, $\mathcal{O}^{\rm CPL}$ and $\mathcal{O}^{\varphi}$ are the observables computed with the CPL parameterization or the dark energy model respectively, the factors $\mathcal{O}^{\rm data} / \mathcal{O}^{\varphi}$ ensure that we fit the observables using the data measurements' relative errors, and ``Cov'' represents the associated covariance. Here, we work with the combined data and minimize $\chi^2_{\rm tot} = \chi^2_{\rm DESI} + \chi^2_{\rm SNe} + \chi^2_{\rm CMB}$. Essentially, this determines the best fit representation for the dark energy model in terms of $(w_0, w_a)$ parameters given the observables predicted by the dark energy model and the associated uncertainties of the observations at the redshifts that have been probed in
the relevant surveys.

\begin{figure}[htbp]
    \centering
    {%
        \includegraphics[width=\columnwidth]{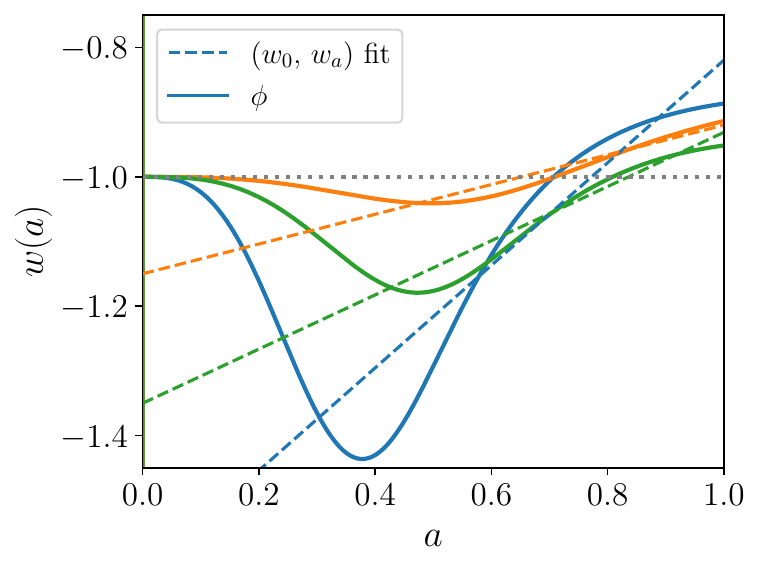}
    }
    \hfill
    {%
        \includegraphics[width=\columnwidth]{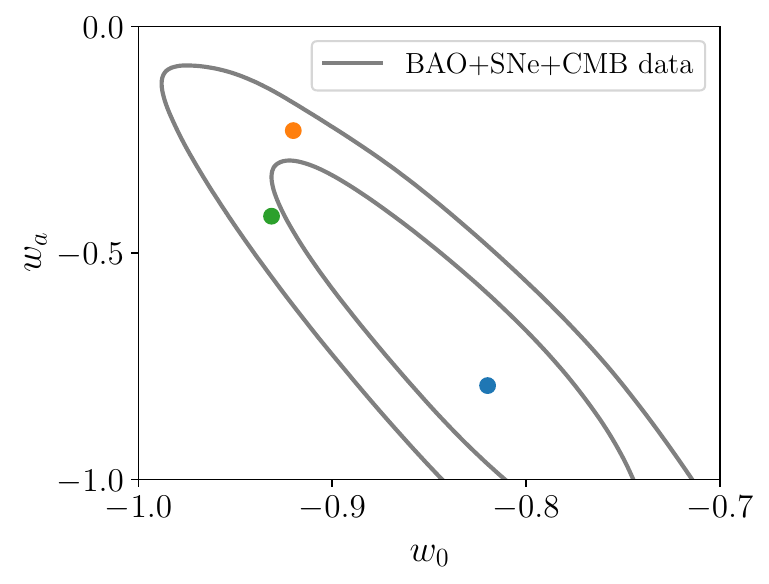}
    }
    \caption{Top: equation of state, $w(a)$, for a representative modified gravity models from \cref{eq:fullaction} alongside the corresponding best fit CPL models when fitting directly to the BAO, SNe, and CMB data (Eq.~\ref{chi2}). Bottom: projection of these $(w_0, w_a)$ fits onto latest data constraints \cite{DESI:2024mwx}.}\label{Fig:fit_model_b}
\end{figure}

It is useful to illustrate this for a few representative models as can be seen in \cref{Fig:fit_model_b}. Consider the $w(a)$ that arise from the theory given by \cref{eq:fullaction} for particular choices of $(\xi, \lambda, V_0, V_1)$ and for a cosmology with $\Omega_m = .31$ and $H_0 = 69$\,km\,s$^{-1}$\,Mpc$^{-1}$. Then, for each value of these parameters, we use \cref{eq:hfit} to find the corresponding values of $(w_0, w_a)$. The blue line is given by $(\xi, \lambda, V_0, V_1)=(2.30, 0.75, 0.94, 5.29)$ and $(w_0, w_a)=(-0.82, -0.79)$, the yellow line is given $(1.50, 0.82, 0.65, 2.67)$ and $(-0.92, -0.23)$, and the green line is given by $(2.10, 1.42, 0.72, 2.24)$ and $(-0.93, -0.42)$. These models cross the phantom divide quite recently ($z<1$), but they differ quite substantially in terms of how far they delve into the phantom regime, as well as how rapidly they subsequently thaw. It is precisely this dynamical flexibility that allows these representative models to map a large region of the $(w_0, w_a)$ plane. 

\begin{figure*}[htbp]
    \centering
    \begin{minipage}[b]{0.9\textwidth}
        \centering
        \includegraphics[width=\columnwidth]{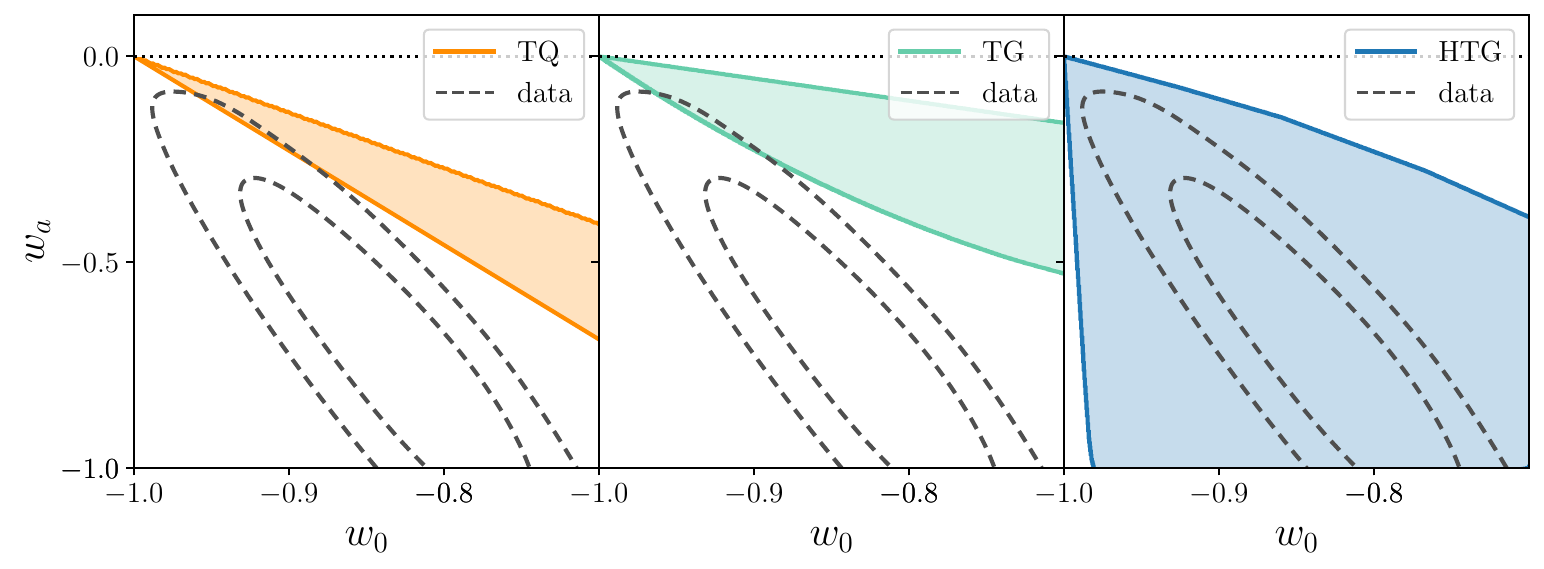}
        \label{fig:subfig1}
    \end{minipage}
    \caption{68\% and 95\% C.L. posterior distribution of the CPL parameters (dashed line) from the combined BAO+CMB+SNe data, where we use the data sets coming from  DESI \cite{DESI:2024kob}, Pantheon+ \cite{Scolnic:2021amr}, and Planck+ACT lensing \cite{Planck:2018vyg, Planck:2019nip,ACT:2023kun, ACT:2023dou}. Overlayed are the CPL parameters, obtained by fitting \cref{chi2}, for thawing quintessence (TQ), thawing gravity (TG), and hilltop thawing gravity (HTG).}\label{Fig:CPL_fit}
\end{figure*}

To see how extensively we can cover the $(w_0, w_a)$ plane we sample $\simeq$ 100,000 models from the HTG model given in \cref{eq:fullaction} on a Latin hypercube in the ranges $\xi \in [0.0, 3.5]$, $\lambda \in [0.0, 3.5]$, and $V_0 \in [0.0, 1.0]$, and for the cosmological parameters $\Omega_{\rm m}\in [0.28, 0.34]$ and $h \in [0.66, 0.72]$. Additionally, as the data strongly favors models for which $w_0 > -1$ (the fitted value of the equation of state today), we will restrict ourselves to models which have successfully crossed the phantom divide and have $w_0 > -1$. As both the model proposed here and the TG model of \cite{Ye:2024ywg} are similar in that they allow for a recent phantom crossing, it will be instructive to compare them. To do this, we also repeat the sampling procedure $\xi \in [0.0, 3.5]$ and $\lambda \in [0.0, 3.5]$ for the TG model (recall that it has one less parameter). To complete the comparison, we also can perform an analogous scan across the {\it minimally coupled} TQ model; in this case the model essentially maps onto the parametrized model explored in detail in \citet{Wolf:2024eph}. 

As depicted in \cref{Fig:CPL_fit}, the model in Eq.~(\ref{eq:fullaction}) saturates the posterior distribution of the combined BAO+SNe+CMB data and can be mapped across the $(w_0, w_a)$ plane when directly fitting the data (c.f.~\cite[Fig.~3]{Wolf:2024eph} whereas the minimally coupled TQ model asymptotes to a particular region of the $(w_0, w_a)$ plane and shares only a small overlap with the posteriors constraints). We also find, similarly to \cite{Ye:2024ywg}, that the TG model occupies a small slice of viable parameter space. However, the fitting procedure used here pushes these models further away from the posterior constraints. This is because, when determining the $(w_0, w_a)$ representation for their models, \cite{Ye:2024ywg} directly fit the equation of state $w(a)$ over a range of recent redshifts ($z < 1$). The fitting procedure here uses the precise redshifts and uncertainties from current BAO and SNe constraints and, crucially, also includes CMB data from $z\simeq1100$. As can be seen from looking at the posteriors for various combinations of data sets in \cite{DESI:2024mwx}, including the CMB will naturally pull the best fit CPL parameters into the upper right region of the $(w_0, w_a)$ plane. 

Thus, even given that this fitting procedure takes account for all the data across these epochs, the model given by \cref{eq:fullaction} has dynamics that enables it to occupy the lower left regions of the posterior constraints. This comes from the fact that the model given by \cref{eq:fullaction}, when compared to its cousin in \cite{Ye:2024ywg}, can move further into the phantom region and evolve more rapidly, which means that the corresponding, derived, $(w_0, w_a)$ values can occupy the lower left region favored by the data. 

Does this mean current data is favouring the HTG model? To answer this question, we turn to a direct comparison between the observables generated by the model and the full (uncompressed) data, bypassing the step of estimating the CPL parameters. We use the likelihoods as implemented in \texttt{Cobaya} \cite{Cobaya, Cobaya2} for
the same data as was used in the DESI analysis \cite{DESI:2024mwx}; i.e., DESI 2024 BAO measurements \cite{DESI:2024mwx, DESI:2024uvr, DESI:2024lzq}, the SNe Ia Pantheon+ samples \cite{Brout:2022vxf, Scolnic:2021amr}, the official \texttt{plik} likelihood for TTTEEE CMB measurements from Planck 2018 \cite{Planck:2019nip, Planck:2018vyg},  and the CMB lensing from ACT DR6 \cite{ACT:2023dou, ACT:2023kun}. 

The biggest difference between the compressed data used in the analysis above and what we will do now is that the CMB data was effectively compressed into just two numbers that can be computed from background quantities. However, the full Planck data consists of the low-$\ell$ power spectra in the range $2 \leq \ell \leq 29$, the high-$\ell$ power spectra covering $30 \leq \ell \leq 2508$ for the temperature auto-correlation ($TT$), and $30 \leq \ell \leq 1996$ for the $TE$ and $EE$ components \cite{Planck:2019nip}. Likewise, CMB lensing from ACT DR6 consists of the ACT CMB reconstructed lensing power spectra between the scales $40 < L < 763$  \cite{ACT:2023dou, ACT:2023kun}. In order to compare this HTG model directly with the combined CMB data, we use \texttt{hi\_class} to compute the full linear cosmological perturbations and CMB power spectra, and model the non-linear matter power spectrum with the \textsc{halofit} algorithm \cite{halofit1, halofit2}.

\begin{figure}
    \centering
    \includegraphics[width=\columnwidth]{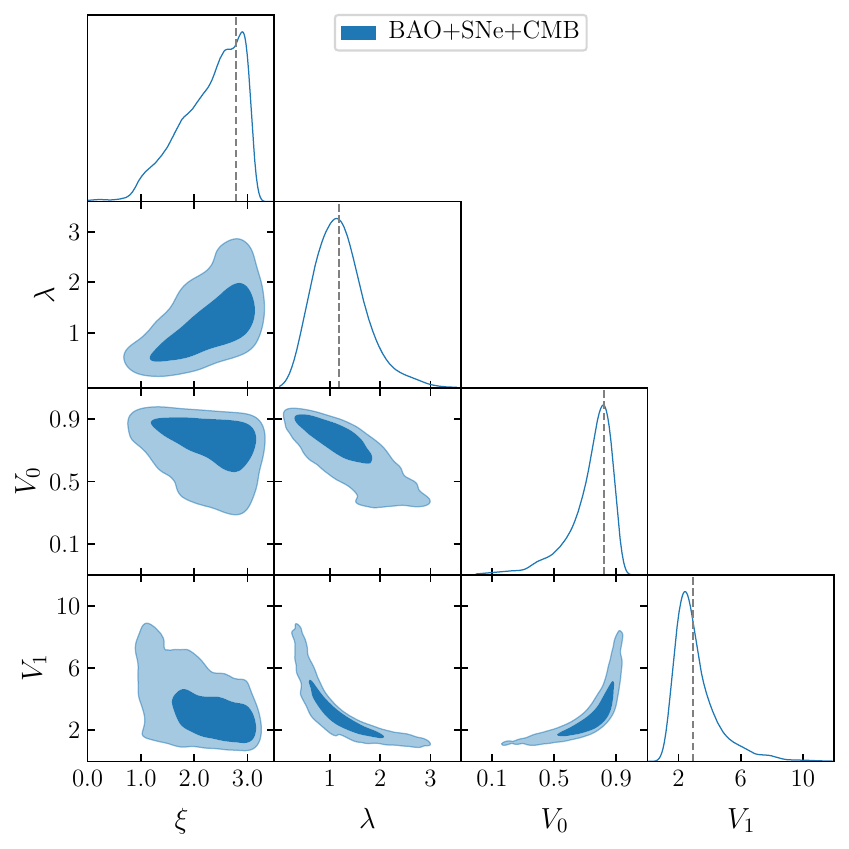}
    \caption{68\% and 95\% C.L. posterior distributions for HTG from the combination of DESI BAO data \cite{DESI:2024mwx}, Pantheon+ SNe data \cite{Scolnic:2021amr}, and CMB data \cite{Planck:2018vyg, Planck:2019nip,ACT:2023kun, ACT:2023dou} obtained from sampling uniformly in the parameters given in Table~\ref{priors}. The dashed gray line gives the best fit parameters from the MCMC.}
    \label{fig:mcmc_negexp}
\end{figure}

\begin{figure}
    \centering
    \includegraphics[width=\columnwidth]{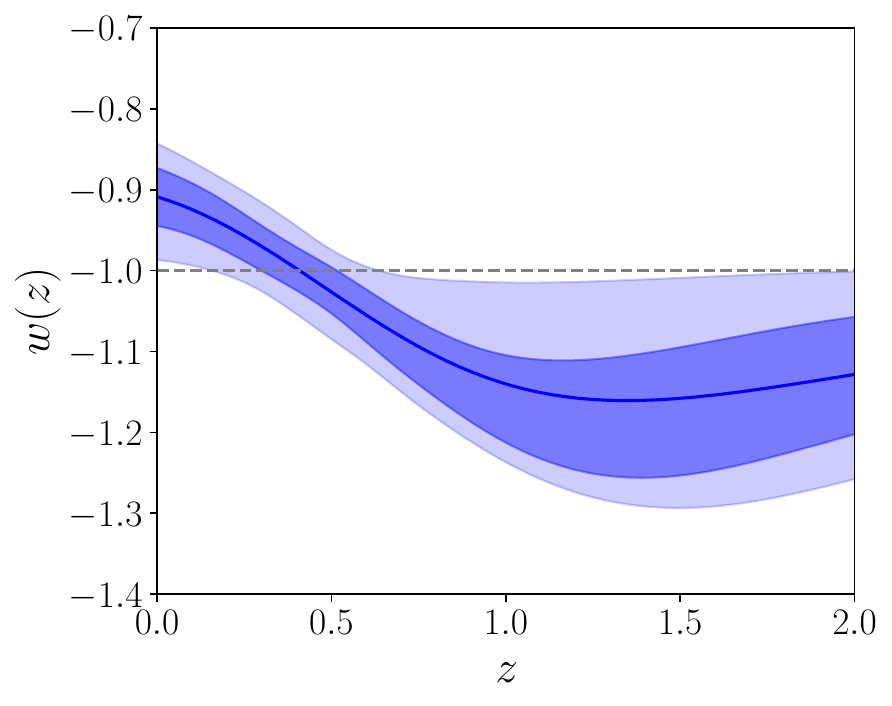}
    \caption{Reconstruction of the equation of state $w(z)$ (1- and 2-$\sigma$ levels) between $0 < z < 2$ from the posteriors obtained in \cref{fig:mcmc_negexp}. The equation of state is phantom, and then sharply evolves at recent times to cross the phantom divide and favors $w_0 > -1$ today.}
    \label{fig:wzconstraint}
\end{figure}

We now obtain constraints on the model parameters $(\xi, \lambda, V_0, V_1)$, exploring the parameter space using a Metropolis-Hasting
algorithm \cite{metropolismc, hastingsmc, Lewis:2002ah, Lewis:2013hha} by
sampling the parameters in Table~\ref{priors} and imposing uniform priors. We adopt a Gelman-Rubin convergence criterion \cite{gelmanrubin} of $R-1 < 0.03$. One can see the posterior distributions of these parameters in \cref{fig:mcmc_negexp} and the reconstructed equation of state $w(z)$ between $0 < z < 2$ is given in \cref{fig:wzconstraint}.

\begin{table}[h!]
\centering
\begin{tabular}{|c|c|}
\hline
\multicolumn{2}{|c|}{\textbf{Cosmological Parameters}} \\
\hline
$\omega_b$ & $\mathcal{U}[0.005, 0.04]$ \\
$\Omega_m$ & $\mathcal{U}[0.2, 0.7]$ \\
$H_0 \ [\text{km/s/Mpc}]$ & $\mathcal{U}[40, 90]$ \\
$n_s$ & $\mathcal{U}[0.9, 1.1]$ \\
$\ln 10^{10} A_s$ & $\mathcal{U}[2.0, 4.0]$ \\
$\tau$ & $\mathcal{U}[0.01, 0.2]$ \\
\hline
\multicolumn{2}{|c|}{\textbf{Dark Energy Parameters}} \\
\hline
$\xi$ & $\mathcal{U}[0, 3.5]$ \\
$\lambda$ & $\mathcal{U}[0, 3.5]$ \\
$V_0$ & $\mathcal{U}[0, 1.0]$ \\
\hline
\end{tabular}
\caption{Prior distributions used in the cosmological parameter inference for the dark energy model given by \cref{eq:fullaction}. $\mathcal{U}(a, b)$ stands for a uniform distribution in the range $[a, b]$.}\label{priors}
\end{table}

For the purposes of this letter, we are particularly focused on whether nonminimal models are, in any way favoured relative to minimally coupled quintessence or even $\Lambda$CDM. To do so, we just need to find the model, given by \cref{eq:fullaction}, that maximizes the likelihood, ${\cal L}$ (or, alternatively, minimizes $\chi^2=-2\ln {\hat{\cal L}}$)\footnote{We use the lite version of Planck likelihood in the minimization.} for this data-set which consists of the same BAO, SNe, and CMB data used in the recent DESI results \cite{DESI:2024mwx}.

For the HTG model in \cref{eq:fullaction}, we find $\chi^2_{\rm HTG} \simeq 2424$, whereas for the TG model \cite{Ye:2024ywg} we find $\chi^2_{\rm TG} \simeq 2432.7$, leading to a substantial $\Delta \chi^2 \simeq 8.7$ between the two models. The TQ model has $\chi^2_{\rm TQ} \simeq 2434$, $\Lambda$CDM has $\chi^2_{\rm \Lambda CDM} \simeq 2436.8$, and $w_0 w_a$CDM has $\chi^2_{\rm CPL} \simeq 2430$. Thus the TG model of \cite{Ye:2024ywg} and the TQ model of \cite{Wolf:2024eph, Wolf:2023uno} perform similarly in terms of their ability to fit the data, with a $\Delta \chi^2 \simeq 1.3$.
 
With a $\Delta \chi^2 \simeq 12.8$, the HTG model seems to be strongly preferred over $\Lambda$CDM (and with $\Delta \chi^2 \simeq 6$, it is even better than $w_0 w_a$CDM). Yet, it is useful to refine this comparison by looking at a more nuanced information criteria that accounts for the number of parameters. Here, we consider the Akaike Information Criterion (AIC) \cite{Liddle:2007fy, AIC_1974, BIC_1978}, which is given by
$
 {\rm AIC}=2k+\chi^2    
$,
where $k$ is the number of parameters in the model. Regarding the AIC, $\Lambda$CDM has one free parameter given by $\{\Lambda\}$, whereas the nonminimal HTG model has five free parameters given by $\{V_1, V_0, \xi, \lambda, {\dot \varphi}_{\rm ini}\}$. Thus, $\Delta k \equiv k^{\rm \Lambda CDM} - k^{\rm HTG} = -4$ and $\Delta {\rm AIC}\simeq 4.8$, which still favors the HTG model over the $\Lambda$CDM model even when accounting for introducing more parameters; however, the evidence is far less substantial than the initial $\chi^2$ would naively suggest.\footnote{One might also want to consider the Bayesian Information Criterion (BIC), given by
$
    {\rm BIC}=k\ln(n)+\chi^2 
$,
where $n\simeq 2400$ is the number of data points. Using the BIC to compare $\Lambda$CDM and HGT, we have $\Delta {\rm BIC}\simeq -18.3$, which strongly favors $\Lambda$CDM. However, one may be worried about using the BIC here as the posteriors display some non-Gaussianity, meaning that Laplace's approximation (which BIC relies upon) breaks down.} When comparing with the $w_0 w_a$CDM model, we have $\Delta k \equiv k^{w_0w_a\rm CDM} - k^{\rm HTG} = -3$, leading to $\Delta {\rm AIC}\simeq 0$, indicating that they are effectively indistinguishable statistically.

\textit{Discussion.---}
In  this letter, we have shown that it is possible to construct a model of dark energy (extending the proposal of \cite{Ye:2024ywg}) that can span a large area of the ($w_0,w_a$) plane, easily covering the region constrained by current observations. A crucial ingredient is that the scalar field is nonminimally coupled to gravity and one might interpret what we found as evidence for new physics. We show that is not the case. While the new model does substantially improve the goodness of fit (or the $\chi^2=-2\ln {\hat{\cal L}}$) relative to $\Lambda$CDM (as well as to $w_0 w_a$CDM), if one takes into account the number of free parameters required to do so, one finds that, as in the discussion of minimally coupled quintessence in \cite{Wolf:2024eph}, the evidence is inconclusive. 

While the evidence for this model is inconclusive, it is, nevertheless, interesting that a nonminimal coupling plays such an important role. What would it take to strengthen the evidence? A first step would be to reduce the number of free parameters in the model. We now have five free parameters and the bare minimum we can have, for a scalar field model, is three (one for the nonminimal coupling, one for a monomial potential, and one for the initial conditions); even in that case, for the same value of the $\chi^2$ the evidence would not be conclusive. The only way the evidence might be strengthened is with future, more constraining data, reinforcing what has been found with the current data, but at higher significance.
An interesting consequence of having a nonminimal coupling is that it might lead to other effects related to
structure formation \cite{Ferreira:2019xrr, Ruiz:2014hma, Alonso:2016suf, Wen:2023bcj, Ruiz-Zapatero:2022xbv, Baker:2013hia, Ferreira:2010sz}, gravitational waves \cite{Baker:2017hug, Ezquiaga:2018btd, Dalang:2019rke, Wolf:2019hun, Ezquiaga:2017ekz} and, most significantly, fifth force tests \cite{Burrage:2017qrf, Joyce:2014kja}. We leave for future work the study of the impact of the nonminimal coupling at small scales and whether the effects can be screened to satisfy the local tests of Gravity (see \cite{Ye:2024zpk} for some recent discussion of this in the context of the TG model that answers this question in the affirmative). However, it is important to remind ourselves that such signatures will impose very tight, independent, constraints on these types of models.

\textit{Acknowledgements.---}
We are very grateful to Gen Ye and Alessandra Silvestri for helpful discussions. 
We acknowledge the \texttt{hi\_class} developers, Emilio Bellini, Miguel Zumalac\'arregui and Ignacy Sawicki, for sharing a private version of the code.
WJW is supported by the HAPP Centre at St.~Cross College, University of Oxford. CGG is supported by the Beecroft Trust.
PGF is supported by STFC and the Beecroft Trust.

\bibliography{refs}

\end{document}